# VARIABILITY MODELING FOR CUSTOMIZABLE SAAS APPLICATIONS

Ashraf A. Shahin[1, 2]

[1]College of Computer and Information Sciences,
Al Imam Mohammad Ibn Saud Islamic University (IMSIU)
Riyadh, Kingdom of Saudi Arabia
ashraf_shahen@ccis.imamu.edu.sa
[2]Department of Computer and Information Sciences,
Institute of Statistical Studies & Research, Cairo University, Egypt

## ABSTRACT

*Most of current Software-as-a-Service (SaaS) applications are developed as customizable service-oriented applications that serve a large number of tenants (users) by one application instance. The current rapid evolution of SaaS applications increases the demand to study the commonality and variability in software product lines that produce customizable SaaS applications. During runtime, Customizability is required to achieve different tenants' requirements. During the development process, defining and realizing commonalty and variability in SaaS applications' families is required to develop reusable, flexible, and customizable SaaS applications at lower costs, in shorter time, and with higher quality. In this paper, Orthogonal Variability Model (OVM) is used to model variability in a separated model, which is used to generate simple and understandable customization model. Additionally, Service oriented architecture Modeling Language (SoaML) is extended to define and realize commonalty and variability during the development of SaaS applications.*

## 1. INTRODUCTION

Software as a Service (SaaS) is a cloud computing service model in which applications are delivered to customers as a service. Instead of developing and maintaining a version of application code for each individual tenant, SaaS application serves thousands of tenants with one single application instance [1].

One of the most prominent approaches for achieving the big variation between tenants' requirements is providing an application template with unspecified parts that can be customized by selecting components from a provided set of components. These unspecified parts are called customization points of an application [8].

SaaS application vendors provide customization tools to allow tenants to customize the running SaaS application instance without stopping or restarting it. During run-time, components are associated and disassociated to/from customization points based on each tenant's requirements.

Developing such applications and providing tenants with easy and understandable customization model are very complex tasks due to the large number of customizations with very complicated relationships. Furthermore, we must consider how to improve flexibility and reusability during the development of SaaS applications to facilitate quick reactions to business changes.

Software product line (SPL) engineering is one of the prominent methodologies for developing software product families in a reusable way. The main idea of SPL engineering is to define and realize commonality and variability of software product line. The variability of the software product line is exploited to derive applications tailored to the specific needs of different customers [2].

However, not all variation points that are specified during the development processes can be provided as customization points during run-time. For example, SaaS developer can define variation point to specify how to isolate tenants' data, which cannot be provided as customization point. Additionally, to provide a customization point, we have to provide a technique to associate and disassociate variants to/from customization point during run-time without stopping or restarting the running SaaS application instance, which is not essential for the variation points during the development time.

Based on the previous SaaS applications research literature [3, 4, 5, 6], we can specify two challenges: the first one is how to define and realize commonalty and variability during the development process. The second issue is how to provide tenants with simple and understandable customization model. Although, there is a strong relationship between the commonality and variability in SaaS applications product lines and the provided customization model, most of current researches focus only on one of these challenges.

In this paper, Orthogonal Variability Model (OVM) is used to model variability in a separated model from other models. Variability model is exploited to generate customization model. Service oriented architecture Modeling Language (SoaML) is extended to realize variability model during the development of SaaS applications.

The rest of the paper is structured as follows. Section 2 gives some background materials. Section 3 gives a short overview of related works. Section 4 presents our approach to model variability in customizable SaaS applications. Section 5 outlines the Implementation of the proposed approach. Section 6 concludes the paper.

## 2. BACKGROUND

### 2.1. Orthogonal Variability Modelling (OVM)

OVM is a proposal for documenting software product line variability [2]. In OVM, only the variability of the product line is documented. In this model, a *variation point* (*VP*) documents a variable item and a *variant* (*V*) documents the possible instances of a variable item. All VPs are related to at least one V and each V is related to at least one VP. Each VP is either internal VP or external VP. Internal VP has associated variants that are only visible to developers but not to customers. The external VP has associated variants that are visible to developers and customers. Figure 1, taken from [2], shows the graphical notation used in OVM.

In OVM, optional variants may be grouped in alternative choices. This group is associated to a cardinality [*min…max*]. Cardinality determines how many Vs may be chosen in an alternative choice, at least *min* and at most *max* Vs of the group.

In OVM, constraints between nodes are defined graphically. A constraint may be defined between Vs, VPs, and Vs and VPs. A constraint may be either *excludes* or *requires* constraint. An *excludes* constraint specifies a mutual exclusion, for instance, a variant excludes a VP means that if the variant is chosen to a specific product then the VP must not be bound, and vice versa. A *requires* constraint specifies an implication, for instance, a variant requires a VP means that always the variant is part of the product, and the VP must be also part of that product. Figure 2 depicts a simple example of an OVM inspired by the travel agent industry.

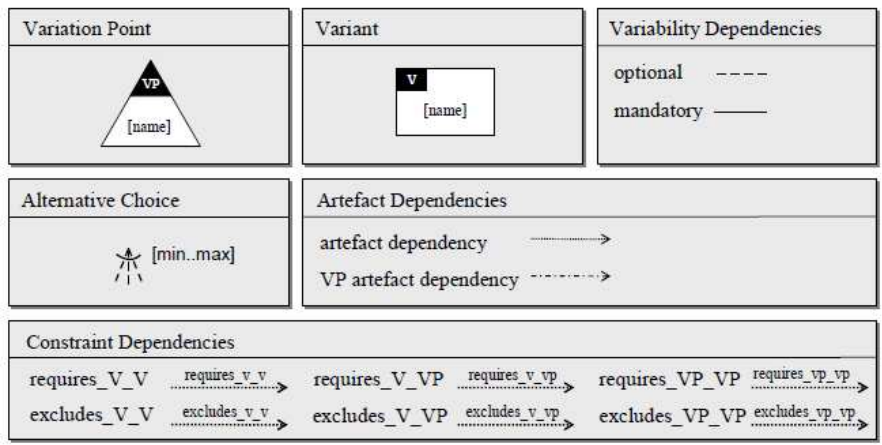

Figure 1. Graphical notation for OVM [2]

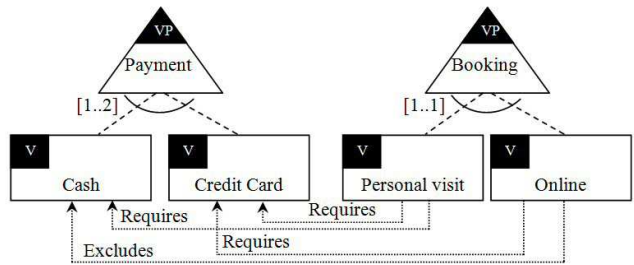

Figure 2. OVM Example: Travel agency industry

## 2.2. SoaML

SoaML is a UML profile extension introduced by OMG to provide a standard way to architect and model SOA solutions. In SoaML, a service can be specified using three approaches [7]: *simple interface*, *service interface* and *service contract*. *Simple interface* defines one-way service that does not require a protocol. *Service interface* defines bi-directional services. *Service contract* defines the agreement between participants to provide and consume the service. The logical capabilities of the services are represented as *capabilities*. *Capabilities* identify or specify functions or resources that are offered by services.

People, organizations, or components that provide or use services are represented as *participants*. *Participants* provide or consume services via ports. *Participants* may provide any number of services and may consume any number of services. In SoaML, *Services Architecture diagram* describes the roles played by *Participants*, their responsibilities, and how they work together to meet some objectives using services [7].

## 3. RELATED WORK

As mentioned earlier in the introduction, there are two challenges to be considered: the first one is how to define and realize commonalty and variability during the development process. The second is how to provide the tenants with simple and understandable customization model. Many researches have been done to address these challenges [8, 9, 10]. However, most of these researches focus only on one of these challenges.

For commonalty and variability modeling during the development process, in [11], Chakir et al. extended SoaML profile to model functional and non-functional variability in the domain services. In [10], Park et al. modeled commonality and variability in service-oriented applications at two levels: composition level and specification level. At the composition level, the authors extended UML Activity Diagram to describe variability in the flow of domain services that fulfill business processes. At the specification level, meta-model was proposed to describe the variability in the domain-services (e.g., properties, operations, messages). In [12, 13], Abu-Matar et al. introduced a multiple view SOA variability model based on feature modeling to model variability during the development time. The authors used UML and extended SoaML to describe the variability model.

For customization modeling, Tsai et al. in [14] modeled customizations using Orthogonal Variability Model and guided the tenants during the customization process based on mining existing tenants' customization. Lizhen et al. in [8] modeled customizations using metagraph and provided an algorithm to validate customizations performed by tenants. In [15], Tsai et al. proposed ontology-based intelligent customization framework to model customizations and to guide tenants.

In [1], Mietzner et al. modeled variability and customizability of SaaS applications using OVM. They used internal variability to represent variability that is visible to the developers of the product line, and used the external variability to represent variability that is visible to the tenants. This approach is quite similar to our approach in this point. However, the authors associated variants with internal variation points based on customizations selected by the tenants, which is difficult to apply in SaaS applications. This difficulty comes because developers associate variants with internal variation points during development time and each running SaaS application instance serves thousands of tenants at the same time, therefore we cannot stop or restart the running SaaS application instance to apply tenants' customizations.

Moens et al. in [9] proposed approach quite similar to the proposed approach in [1], except they used feature model instead of OVM to model variability. In the feature model, each variation was represented as a feature and realized by a service. SaaS applications are composed and deployed based on features specified by the clients. The authors mapped feature model to code module. We on the other hand mapped variability model to the SoaML model without regard for how a particular variation might be implemented. Because, when we model variations, web-services that realize these variations may not yet be known.

## 4. THE PROPOSED VARIABILITY MODELLING APPROACH

Most of current SaaS applications are developed using service oriented architecture (SOA) model [1, 14], which comprises of five layers:

- *Presentation layer:* contains interface programs that are used to interface SOA applications.
- *Business processes layer:* contains business processes that realize interfaces in *presentation layer*. Business process is composed of services and of other business processes. Activity diagram is commonly used to model business process workflow, which is represented as a sequence of activities.
- *Business services layer:* contains services that perform business processes' activities. Each service can be atomic or composed of other services. Services are specified by *simple interfaces*, *service interfaces,* or *service contracts.*
- *Components layer:* contains participants that provide services (or realize service interfaces) in the *business services layer*. As shown in figure 3, Participant may provide any number of services, and each service may be provided by any number of participants.
- *Operational systems layer:* contains programs and data that are required to provide the service functionalities in the *business services layer*, and contains infrastructure programs (e.g., operating systems, database management systems, runtime environments) that are needed to support SOA applications.

Variants and customization points can be located in any layer of these layers. In our approach, we concerned only with variability and customizability modeling in three layers: *business processes layer*, *business services layer*, and *components layer*.

In our approach, we use OVM to model variability and customizability in SaaS applications. Developers produce customization model by associating variants with variation points. We extend SoaML profile to represent variants, variation points, customization points, and their dependencies during the development process. We define model-to-model transformation mapping to derive specific customizable SaaS application model.

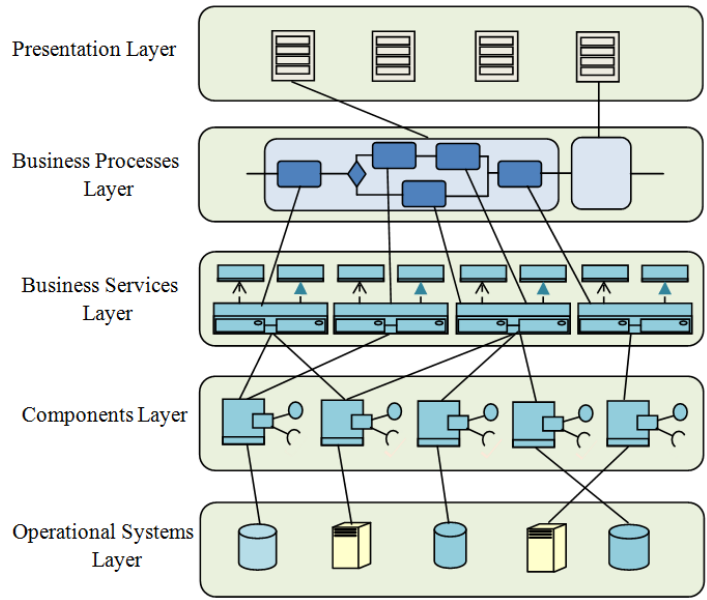

Figure 3. SOA Architecture

### 4.1. Variability Modelling Using OVM

Using OVM, Variability in SaaS applications is modeled in a separated model from other models (e.g. design model). This separation simplifies the variability model and provides understandable customization model to the SaaS tenants.

As mentioned in subsection 2.1, Variation points in OVM have two types: internal and external variation points. We use internal variation points to represent variation points (VPs) in customizable SaaS applications product line, and use the external variation points to represent customization points (CPs) in a customizable SaaS application. Figure 4 shows variability model example using OVM. SaaS developers associate variants (Vs) to internal variation points to produce customizable SaaS applications. Figures 5 and 6 give two different customization models after different associations. Figure 5 shows the customization model driven from the variability model in figure 4 after associating V1 with VP1 and CP3 with VP2. V1 excludes V3, thus V3 was removed. Figure 6 shows the customization model after associating V2 with VP1 and CP2 with VP2 in figure 4. V2 requires V5, therefore v5 was changed from optional to mandatory variant to CP1 and the cardinality of CP1 was changed from [1..2] to [0..1].

Tenants' administrators customize the provided customizable SaaS applications during runtime by associating variants to CPs (external variation points) to produce customized applications for their users. Again, for each customization point, SaaS developers must provide a technique to associate/disassociate variants to/from customization point in the running SaaS application instance during runtime without stopping or restarting the running SaaS application instance.

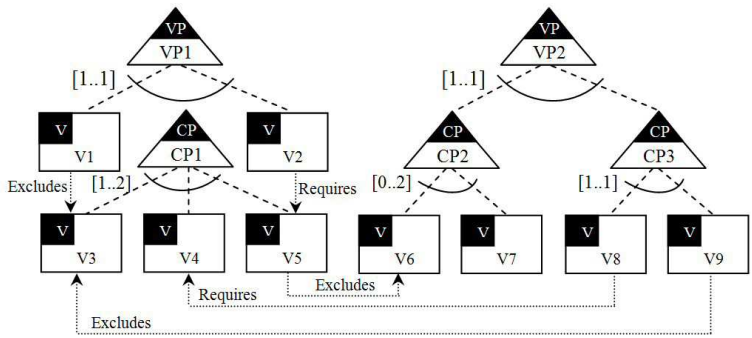

Figure 4. Variability model example using OVM

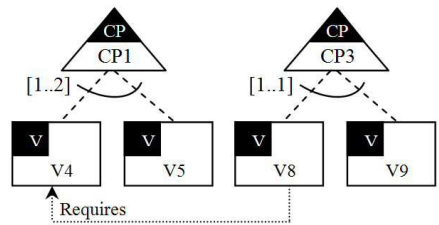

Figure 5. Customization model after associating V1 with VP1 and CP3 with VP2 in figure 4

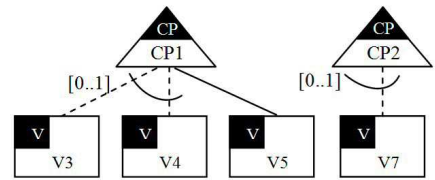

Figure 6. Customization model after associating V2 with VP1 and CP2 with VP2 in figure 4

### 4.2. The Proposed Variability Modelling Profile

To represent variability and customizability during the development process, we extended SoaML profile with the UML profile shown in figure 7. The *AbstractVatiationPoint* stereotype is abstract, is not meant to be used. The stereotype *AbstractVatiationPoint* extends the UML metaclass *Element* with the stereotype attributes *MinAlternativeChoice* and *MaxAlternativeChoice*, which determine how many optional Vs may be chosen in an alternative choice. The stereotypes *CustomizationPoint* and *VariationPoint* are specializations of the stereotype *AbstractVatiationPoint* to represent CPs and VPs, respectively.

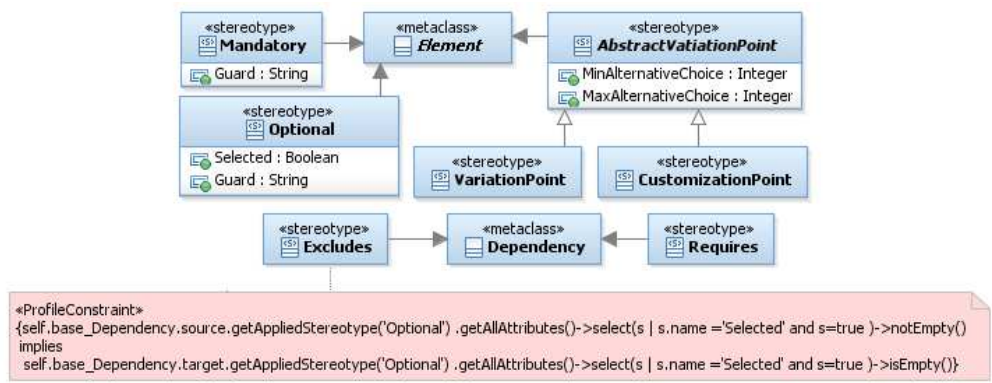

Figure 7. Variability modeling profile

Optional Vs are represented by the stereotype *Optional* with attributes "*Selected*" and *"Guard."* The *"Selected"* attribute will be set to true if the optional V is selected. The *"Guard"* attribute contains condition that must be true in order to traverse V. Mandatory Vs are represented by the stereotype *Mandatory* with the stereotype attribute *"Guard."* Guards contain conditions written by UML Action Language (UAL), which is based on Action Language for Foundational UML (Alf) [16].

The *Excludes* and *Requires* stereotypes extend the UML meta-class *Dependency* to represent constraint dependencies between Vs. Constraint dependencies between Vs are modeled in a separated model to describe dependencies between Vs from different layers. For each stereotype, constraint is defined using Object Constraint Language (OCL) [17]. Constraints will be evaluated in the model, where the stereotype is applied. Figure 7 shows the *Excludes* stereotype constraint.

### 4.3. Modeling Variability in Business Processes Layer

Business process workflow is a sequence of activities, which is modeled using activity diagram. In the activity diagram, the flow of execution is modeled as *activity nodes* connected by *activity edges*. Activity nodes can be *Action*, *ObjectNode*, or *ControlNode*. *Action* represents a single atomic step within activity. *ObjectNode* is used to define object flow in an activity. *ControlNode* is used to coordinate the flows between other nodes.

In the business processes layer, *StructuredActivityNode* is used to defined Vs, CPs, and VPs. A structured activity node is an executable activity node that contains a structured portion of the activity that is not shared with any other structured node. Using *StructuredActivityNode*, we can describe variants' behaviors in SOA SaaS applications, which are loosely coupled web services. Figure 8 shows the *StructuredActivityNode* meta-class in activity diagram meta-model.

Vs that are offered for a particular VP are represented as structured activity nodes inside its VP node. VPs nodes contain UML decision node with outgoing edges to Vs nodes. Vs' guards are used as guards for the decision node outgoing edges. UML merge node is used to bring together the outgoing flows from different alternative Vs to a single outgoing flow.

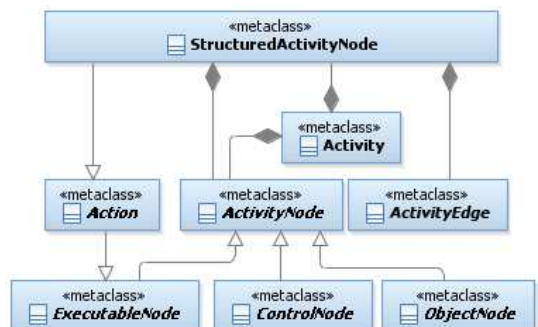
Figure 8. StructuredActivityNode Metaclass in Activity Diagram Meta-model

After representing Vs, VPs, CPs, and their relationships, SaaS developers associate Vs with VPs to produce customizable business processes by exploring the activity diagram and modifying the attribute "*Selected*." Finally, the produced activity diagram is transferred to another activity diagram after applying the associations made by developers. Figure 9 shows an example for transforming VPs.

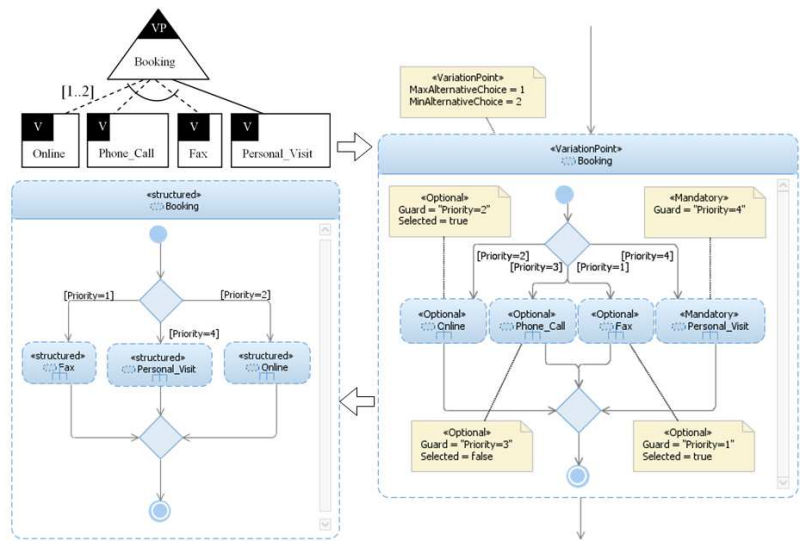

Figure 9. Example of VP representation and transformation in activity diagram

### 4.4. Modelling Variability in Business Services Layer

In the business services layer, service interfaces are defined for services that are required to perform activities in business process layer. For each activity, developer can specify a set of alternative services that satisfy the activity behavior but their interfaces do not match to the defined interface.

To represent service interface variability in the business services layer, VPs and CPs are represented as SoaML *ServiceInterfaces* with the stereotypes <<*VariationPoint*>> and <<*CustomizationPoint*>>, respectively. CPs and VPs contain SoaML *Request* ports for their Vs. *Request* ports are typed by Vs service interfaces and stereotyped by the stereotype <<*Optional*>> or <<*Mandatory* >> to represent optional or mandatory variants. Optional Vs are associated with CPs or VPs by modifying the "*Selected"* attribute. We have moved the optional and mandatory stereotypes from variant service interfaces to the CPs and VPs ports, because the same variant service interface may be participated in different VPs and CPs with different types (optional or mandatory).

Using activity diagram, the behavior of CPs and VPs are described. Structure activity nodes are used to describe how variants perform activities. Activity diagram will be transformed to another activity diagram (as in business processes layer) to apply Vs association. Figure 10 shows an example for service interfaces variability modeling in business services layer.

### 4.5. Modelling Variability in Components Layer

Service components layer contains components that implements service interfaces in the business services layer. In SoaML, service components are represented as Participants. For each service interface in the business services layer, developer can specify a set of alternative participants that can provide the specified service interface. To represent participant variability in the components layer, VPs and CPs are represented as SoaML *Participants*, which implement the provided service through delegation to one of the variant participants. VPs and CPs participants contain service interface ports to delegate their requests to other participants. Ports are stereotyped by the stereotype <<*Optional* >> or <<*Mandatory* >> to represent optional or mandatory variants. The behavior of the CPs and VPs participants are described by activity diagram.

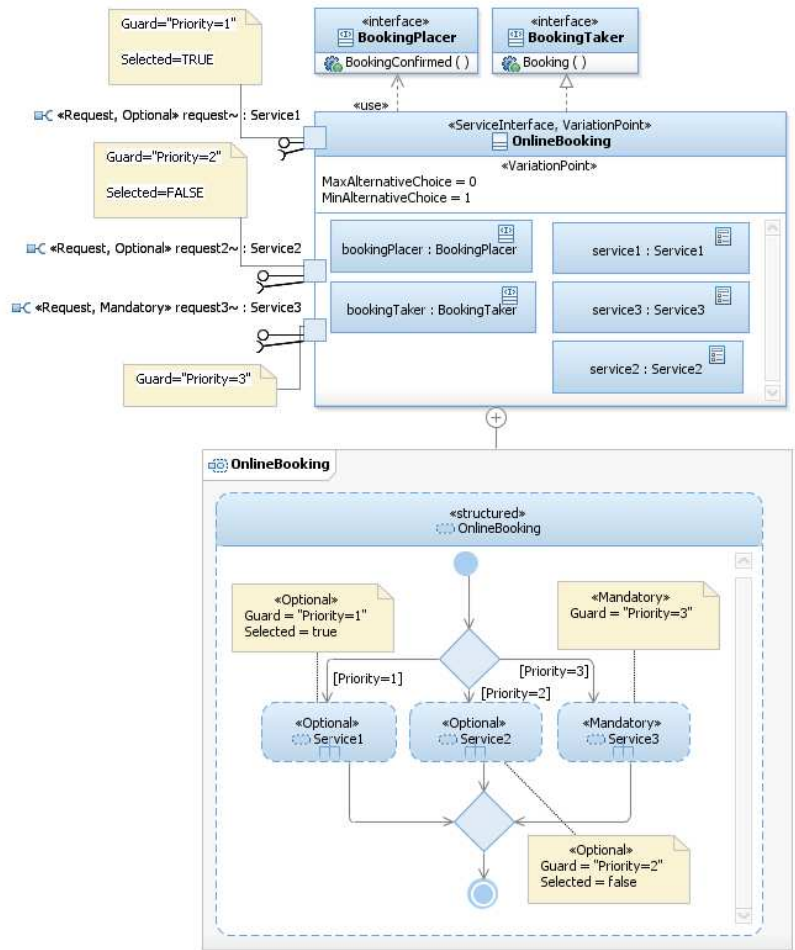

Figure 10. Service interface variability modeling in business services layer

## 5. IMPLEMENTATION OF THE PROPOSED APPROACH

To evaluate the applicability of our approach, we have created the proposed UML profile using Rational Software Architect version 8.5. Additionally, we have created model-to-model transformation prototype based on Eclipse Modeling Framework (EMF). Using Rational Software Architect version 8.5, the prototype has been implemented as an Eclipse plug-in. Mapping model has been created with mapping declarations to specify how to derive a specific customizable SaaS application model.

Finally, we can summarize the contributions of our approach and its prototype as following:

- Modelling variability and customizability of a set of related SaaS applications in a separated model.
- Generating simple customizations models, which will be used to guide tenants during the customization process and to validate customizations made by tenants.
- Extending SoaML profile to represent variability and customizability in different layers of SOA SaaS applications.
- Validating associations of variants with variation points during the development time by evaluating constraints written using OCL.
- Proposing a model-to-model transformation prototype to generate model for a specific application.

However, the proposed approach requires more investigation to develop model-to-model transformation prototype to map variability models represented by OVM to variability models represented by the extended SoaML profile.

## 6. CONCLUSION

In this paper, we have proposed a variability modeling approach to model variability and customizability in customizable SaaS applications. We have applied the concepts of internal and external variability from software product line engineering to distinguish between variability required to develop flexible and reusable SaaS applications and customizability required to customize SaaS applications during run-time.

SoaML profile has been extended to represent variability and customizability for a family of customizable SaaS applications. Once variants are associated with variation points, a model-to-model transformation is performed to generate a specific customizable SaaS application model. OVM has been exploited to model customizability and variability in a separated model, which will be used to generate customization model for a specific SaaS application.

In our future work, we plan to enhance our approach by providing a UML profile to represent OVM models and by developing a model-to-model transformation prototype to transform OVM models to SoaML models and vice versa.

## REFERENCES

[1]  R. Mietzner, A. Metzger, F. Leymann, and K. Pohl, "Variability Modeling to Support Customization and Deployment of Multi-Tenant-Aware Software as a Service Applications," in *2009. PESOS 2009. ICSE Workshop on Principles of Engineering Service Oriented Systems*, 2009, pp. 18–25.

[2]  K. Pohl, G. Böckle, and F. Linden, *Software Product Line Engineering: Foundations, Principles and Techniques*, Springer-Verlag, New York, 2005.

[3]  W.T. Tsai, Y. Huang, and Q. Shao, "EasySaaS: A SaaS Development Framework," in *2011 IEEE International Conference on Service-Oriented Computing and Applications (SOCA)*, Dec 2011, pp. 1–4.

[4]  M. Rosenmüller, N. Siegmund, T. Thüm, and G. Saake, "Multi-Dimensional Variability Modeling," in *Proceedings of the 5th Workshop on Variability Modeling of Software-Intensive Systems*, ser. VaMoS '11. New York, NY, USA: ACM, 2011, pp. 11–20.

[5]  J. Schroeter, S. Cech, S. Götz, C. Wilke, and U. A, "Towards Modeling a Variable Architecture for Multi-Tenant SaaS-Applications," in *Proceedings of the Sixth International Workshop on Variability Modeling of Software-Intensive Systems*, ser. VaMoS '12. New York, NY, USA: ACM, 2012, pp. 111–120.

[6]  J. Kabbedijk, and S. Jansen, "Variability in Multi-tenant Environments: Architectural Design Patterns from Industry," in *Advances in Conceptual Modeling. Recent Developments and New Directions*, ser. Lecture Notes in Computer Science, Springer Berlin Heidelberg, 2011, vol. 6999, pp. 151–160.

[7]  OMG, *Service oriented architecture Modeling Language (SoaML) Specification Version 1.0.1*, Object Management Group, 2012, http://www.omg.org/spec/SoaML/1.0.1/PDF

[8]  C. Lizhen, W. Haiyang, J. Lin, and H. Pu, " Customization Modeling based on Metagraph for Multi-Tenant Applications," in *2010 5th International Conference on Pervasive Computing and Applications (ICPCA)*, 2010, pp. 255–260.

[9]  H. Moens, E. Truyen, S. Walraven, W. Joosen, B. Dhoedt, and F. De Turck, "Developing and Managing Customizable Software as a Service Using Feature Model Conversion," in *2012 IEEE Network Operations and Management Symposium (NOMS)*, 2012, pp. 1295–1302.

[10] J. Park, M. Moon, and K. Yeom, "Variability Modeling to Develop Flexible Service-Oriented Applications," *Journal of Systems Science and Systems Engineering*, 2012, vol. 20, no. 2, pp. 193–216.

[11] B. Chakir, and M. Fred, "Towards a Model Driven Approach for Variability Management in SOA," in *International Conference on Models of Information and Communication Systems (MICS'10), 2010*.

[12] M. Abu-Matar, and H. Gomaa, "Variability Modeling for Service Oriented Product Line Architectures," in *2011 15th International Software Product Line Conference (SPLC)*, Aug 2011, pp. 110–119.

[13] M. Abu-Matar, and H. Gomaa, "Feature-Based Variability Meta-modeling for Service-Oriented Product Lines," in *Models in Software Engineering*, ser. Lecture Notes in Computer Science, Springer Berlin Heidelberg, vol. 7167, 2012, pp. 68–82.

[14] W. Tsai, and X. Sun, "SaaS Multi-Tenant Application Customization," in *2013 IEEE 7th International Symposium on Service Oriented System Engineering (SOSE)*, 2013, pp. 1–12.

[15] W. Tsai, Q. Shao, and W. Li, "OIC: Ontology-Based Intelligent Customization Framework for SaaS," in *2010 IEEE International Conference on Service-Oriented Computing and Applications (SOCA)*, 2010, pp. 1–8.

[16] OMG, *Action Language for Foundational UML, Concrete Syntax for a UML Action Language*, Object Management Group, 2013, http://www.omg.org/spec/ALF/1.0.1/PDF

[17] OMG, *Object Constraint Language (OCL)*, Object Management Group, 2014, http://www.omg.org/spec/OCL/2.4/PDF